\documentclass[fleqn,aps,prl,superscriptaddress,preprint,nobibnotes, reprint,]{revtex4-2}

\usepackage[]{amsmath}
\usepackage{amssymb}
\usepackage[dvips]{graphicx}
\usepackage{color}
\usepackage{tabularx}
\usepackage{mathtools}
\usepackage{algpseudocode}
\usepackage{enumitem}
\usepackage{multirow}
\usepackage{epstopdf}  
\usepackage{bm}
\usepackage{longtable}
\usepackage{booktabs}

\usepackage{makecell}
\usepackage{hyperref}
\usepackage[version=3]{mhchem}
\usepackage{tikz}
\usetikzlibrary{shapes,snakes}
\usetikzlibrary{arrows}
\usepackage{threeparttable}

\usepackage{xparse}

\newcommand{\hmn}[1]{
  \ensuremath{\begingroup\setupHMN #1\endgroup}%
}
\newcommand{\setupHMN}{%
  \doHMN{-}{\HMNoverline}%
  \doHMN{*}{\HMNminverse}%
  \doHMN{i}{\infty}
  \doHMN{_}{\HMNsubscript}
}
\newcommand{\doHMN}[2]{%
  \begingroup\lccode`~=`#1
  \lowercase{\endgroup\let~}#2%
  \mathcode`#1="8000
}
\newcommand{\HMNminverse}[1]{\frac{#1}{m}}
\newcommand{\HMNoverline}[1]{\mkern1mu\overline{\mkern-1mu#1\mkern-1mu}\mkern1mu}
\newcommand{\HMNsubscript}[1]{_{#1}}
\newcommand{\hulllam}{\mathrm{{hull,LAM}}}
\newcommand{\hulldft}{\mathrm{{hull,DFT}}}

\begin{document}
\title{
Discovery of High-Temperature Superconducting Ternary Hydrides via Deep Learning
}

\author{Xiaoyang Wang}
\affiliation{National Key Laboratory of Computational Physics,
  Institute of Applied Physics and Computational Mathematics, Fenghao East Road 2, Beijing 100094, P.R.~China}

\author{Chengqian Zhang}
\affiliation{Academy for Advanced Interdisciplinary Studies, Peking University, Beijing 100871, P. R.~China}

\author{Zhenyu Wang}
\affiliation{{Key Laboratory of Material Simulation Methods \& Software of Ministry of Education and State Key Laboratory of Superhard Materials, College of Physics, Jilin University, Changchun 130012, P.R.~China}}
\affiliation{{International Center of Future Science, Jilin University, Changchun 130012, P.R.~China}}

\author{Hanyu Liu}
\thanks{Corresponding Author: Hanyu Liu, hanyuliu@jlu.edu.cn; Jian Lv, lvjian@jlu.edu.cn; Han Wang, wang\textunderscore han@iapcm.ac.cn; Yanming Ma, mym@jlu.edu.cn}
\affiliation{{Key Laboratory of Material Simulation Methods \& Software of Ministry of Education and State Key Laboratory of Superhard Materials, College of Physics, Jilin University, Changchun 130012, P.R.~China}}
\affiliation{{International Center of Future Science, Jilin University, Changchun 130012, P.R.~China}}

\author{Jian Lv}
\thanks{Corresponding Author: Hanyu Liu, hanyuliu@jlu.edu.cn; Jian Lv, lvjian@jlu.edu.cn; Han Wang, wang\textunderscore han@iapcm.ac.cn; Yanming Ma, mym@jlu.edu.cn}
\affiliation{{Key Laboratory of Material Simulation Methods \& Software of Ministry of Education and State Key Laboratory of Superhard Materials, College of Physics, Jilin University, Changchun 130012, P.R.~China}}

\author{Han Wang}
\thanks{Corresponding Author: Hanyu Liu, hanyuliu@jlu.edu.cn; Jian Lv, lvjian@jlu.edu.cn; Han Wang, wang\textunderscore han@iapcm.ac.cn; Yanming Ma, mym@jlu.edu.cn}
\affiliation{National Key Laboratory of Computational Physics,
  Institute of Applied Physics and Computational Mathematics, Fenghao East Road 2, Beijing 100094, P.R.~China}
\affiliation{HEDPS, CAPT, College of Engineering, Peking University, Beijing 100871, P.R.~China}  

\author{Weinan E}
\affiliation{AI for Science Institute, Beijing 100080, China}
\affiliation{Center for Machine Learning Research, Peking University, Beijing 100871, P.R.~China}
\affiliation{School of Mathematical Sciences, Peking University, Beijing, 100871, P.R.~China}

\author{Yanming Ma}
\thanks{Corresponding Author: Hanyu Liu, hanyuliu@jlu.edu.cn; Jian Lv, lvjian@jlu.edu.cn; Han Wang, wang\textunderscore han@iapcm.ac.cn; Yanming Ma, mym@jlu.edu.cn}
\affiliation{{Key Laboratory of Material Simulation Methods \& Software of Ministry of Education and State Key Laboratory of Superhard Materials, College of Physics, Jilin University, Changchun 130012, P.R.~China}}
\affiliation{{International Center of Future Science, Jilin University, Changchun 130012, P.R.~China}}
\date{\today}

\begin{abstract}

The discovery of novel high-temperature superconductor materials holds transformative potential for a wide array of technological applications.
However, the combinatorially vast chemical and configurational search space poses a significant bottleneck for both experimental and theoretical investigations. 
In this study, we employ the design of high-temperature ternary superhydride superconductors as a representative case to demonstrate how this challenge can be well addressed through a deep-learning-driven theoretical framework. 
This framework integrates high-throughput crystal structure exploration, physics-informed screening, and accurate prediction of superconducting critical temperatures.
Our approach enabled the exploration of approximately 36 million ternary hydride structures across a chemical space of 29 elements, leading to the identification of 144 potential high-\(T_c\) superconductors with predicted \(T_c > 200\) K and superior thermodynamic stability at 200~GPa. 
Among these, 129 compounds spanning 27 novel structural prototypes are reported for the first time, representing a significant expansion of the known structural landscape for hydride superconductors.
This work not only greatly expands the known repertoire of high-\(T_c\) hydride superconductors but also establishes a scalable and efficient methodology for navigating the complex landscape of multinary hydrides.
\end{abstract}

\maketitle
The pursuit of superconductivity at elevated temperatures remains a fundamental and enduring challenge in condensed matter physics. While extensive research has explored a broad spectrum of superconducting materials, including unconventional systems with complex electronic correlations\cite{cupratereview,ironbasedreview,nickelbased}, the recent emergence of compressed hydrides as a prominent class of conventional superconductors with exceptionally high critical temperatures (T$_c$) with above 200 K has garnered considerable attention\cite{SYreview,cuitianreview,lv2020theory}. This surge of interest has been primarily driven by the experimental realization of superconductivity in H$_3$S\cite{H3Sexperiment}, as well as in a series of clathrate binary hydrides such as CaH$_6$\cite{CaH6-exp1,CaH6-exp2}, YH$_6$\cite{YH6-exp1,YH6-exp2}, YH$_9$\cite{YH6-exp2}, and LaH$_{10}$~\cite{LaH10-exp1,LaH10-exp2,LaH10-exp3}, which exhibit T$_c$ values ranging from 215 to 260 K under high-pressure conditions. Notably, these groundbreaking discoveries were inspired by first-principle based structure searching simulations~\cite{CaH6-theoretic,LaH10-theoretic2,LaH10-theoretic,H3S-theoretic1,H3S-theoretic2,YH-theoretic}, underscoring the predictive power of such computational prediction approaches in discovering new materials~\cite{wang2014perspective,oganov2019structure}.

Currently, the exploration of hydride superconductors is progressing beyond traditional binary systems into more intricate ternary systems\cite{SYreview,cuitianreview,lv2020theory}. These complex systems represent a largely uncharted and promising landscape for the identification of new superconducting materials. For instance, there are a variety of predicted or synthesized nonstoichiometric high-T$_c$ ternary alloy hydrides with random metal atom occupation on lattice sites to replicate the structural framework of parent binary hydrides such as CaH$_6$, YH$_9$ or LaH$_{10}$ systems. Notable examples include (La,Y)H$_{10}$ \cite{LaYH-experiment} (\(T_c \approx 253\) K at 183 GPa), (La,Ce)H$_9$ \cite{LaCeH9-experiment} (\(T_c \approx 148-178\) K at 97-172 GPa), and (La,Al)H$_{10}$ \cite{LaAlH-experiment} (\(T_c \approx 223\) K at 164 GPa). In addition, a series of stoichiometric ternary hydrides have been proposed\cite{MXH12,CaY3H24,CaYH12-258, YbLuH,Li2MgH16,LiNaH,sanna2024prediction,LaAcY_Th_H,YSc2H24,YSrH12,RN764,LaMg3H28,LaBeH8-theory,LaThH,LiLaH,LaSc2H24-LHY}, either through high-throughput structural substitutions of known prototype clathrate systems or via extensive crystal structure predictions (CSP). Notable examples include the first stoichiometric high-T$_c$ ternary hydride superconductor, LaBeH$_8$\cite{LaBeH8-theory}, which features a well-defined crystal structure and has been experimentally synthesized\cite{LaBeH8-experiment} with a measured T$_c$ of 110 K at a moderate pressure of 80 GPa. Additionally, theoretical predictions highlight compounds such as Li$_2$MgH$_{16}$\cite{Li2MgH16}, NaLi$_3$H$_{23}$\cite{LiNaH}, Li$_2$NaH$_{17}$\cite{LiNaH}, and LaSc$_2$H$_{24}$\cite{LaSc2H24-LHY}, all of which are anticipated to exhibit T$_c$ values exceeding room temperature under high-pressure conditions.

While the vast configurational space of ternary hydrides offers a promising avenue for advancing the boundaries of high-temperature superconductivity, it also presents considerable challenges for theoretical investigations. The complexity stems from the immense number of possible stoichiometries and structural variations, which must be thoroughly explored to identify stable and superconducting phases. 
Traditional methods, such as high-throughput structural substitutions\cite{MXH12,YbLuH,jiang2025data}, enable the rapid screening of a wide range of compositional spaces but are often limited to existing structural prototypes. 
This limitation also raises concerns about the thermodynamic stability of the proposed structures, 
as an incomplete search of the convex hull may overlook important competing phases. 

In contrast, first-principles-based CSP methods offer a more comprehensive and unbiased approach by exploring the configurational space without relying on existing structural templates\cite{wang2014perspective,oganov2019structure}. 
However, these methods are computationally demanding particularly for ternary or even multinary systems, thus restricting their application to a limited number of chemical systems. \cite{LiLaH,Li2MgH16,LaSc2H24-LHY,LaBeH8-theory,LaMg3H28,LiNaH},
Furthermore, accurate prediction of T$_c$s for candidate structures necessitates computationally intensive electron-phonon coupling (EPC) calculations, adding a further layer of complexity to the theoretical investigation of these superhydrides.  
These challenges underscore the need for the development of theoretical approaches to effectively navigate the complex landscape of ternary hydride superconductivity.

\begin{figure*}
\begin{center}
\includegraphics[width=1.0\textwidth]{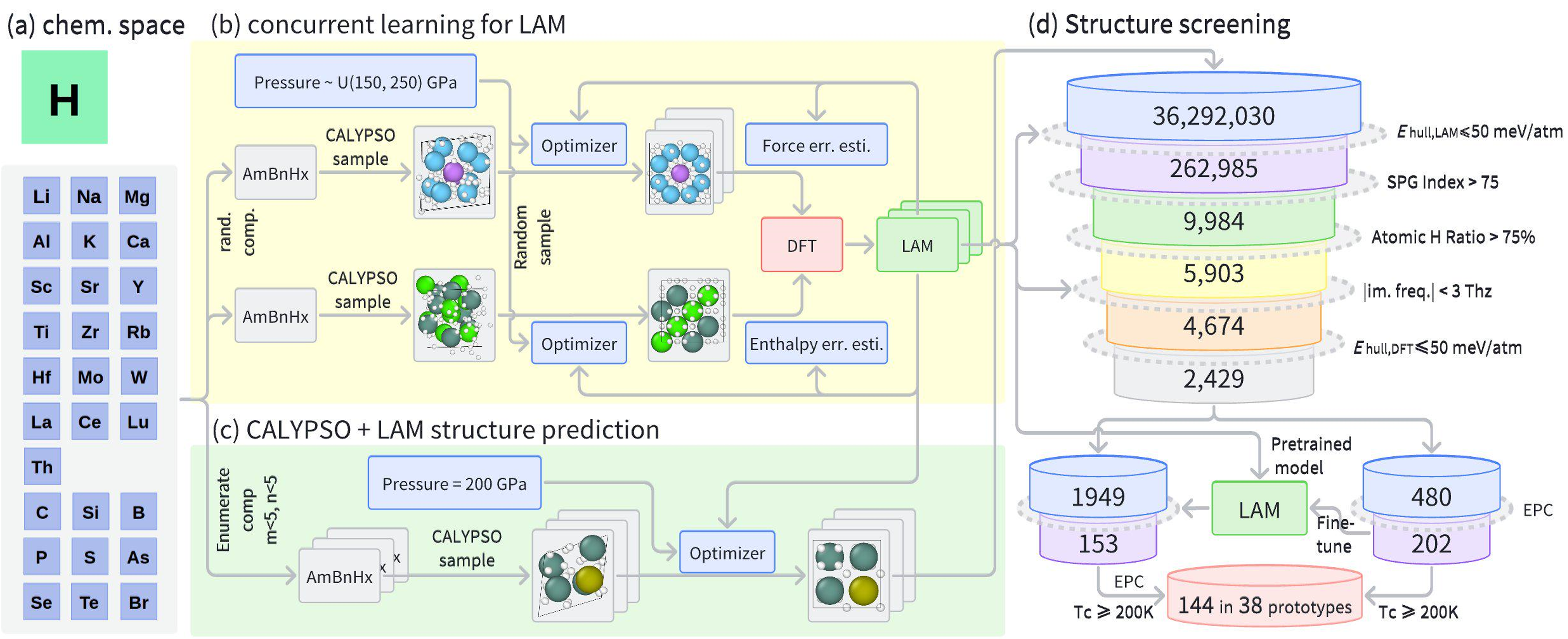}\\[5pt]  
\caption{Schematic representation of the computational framework and workflows utilized in this study. (a) The chemical space covered in this work, comprising 29 elements, including hydrogen, 19 metallic elements, and 9 non-metallic elements. (b) DPA construction based on the concurrent learning scheme\cite{wang2024concurrent}. (c) Structure prediction using the CALYPSO method, with the DPA model employed for energy estimation during the large-scale exploration of ternary hydride structures. (c) Workflow for property screening, outlining the stepwise filtering criteria applied to identify high-temperature superconducting candidates.} 
\label{fig:fig1}
\end{center}
\end{figure*}

To address above mentioned challenges, here we introduce a theoretical framework for high-throughput discovery of high-temperature hydrogen-rich superconductors (Fig.~\ref{fig:fig1}). 
Central to this framework is the Large Atomic Model (LAM), 
a deep learning  model that consistently demonstrates high accuracy in characterizing the enthalpy landscape of ternary hydrides across a chemical space of 29 elements under high pressure.
The framework incorporates the CALYPSO structure prediction method~\cite{CALYPSO,CALYPSO2}, utilizing the LAM as an enthalpy predictor to systematically and comprehensively explore ternary hydrides with favorable enthalpies.
Furthermore, the identified hydride structures undergo a high-throughput screening process,  utilizing a series of filtering criteria and a superconducting transition temperature (T$_c$) model fine-tuned from the LAM, to identify the promising candidates for high-temperature superconductivity. 
Through the current framework, we explored approximately 36 million ternary hydride structures, 
leading to the identification of 144 potential high-temperature hydride superconductors across 38 structural prototypes. 
Notably, 129 of these hydrides, spanning 27 structural prototypes, represent novel predictions that, to the best of the authors' knowledge, have not been previously reported in the literature.

The LAM for compressed hydrides encompasses a chemical space of 29 elements of interest, including hydrogen, 19 metallic elements, and 9 non-metallic elements as illustrated in Fig.~\ref{fig:fig1}(a).
Previous studies have shown that the combination of these non-hydrogen elements with hydrogen may exhibit potential superconductivity~\cite{SYreview,cuitianreview,lv2020theory}.
The model is based on the Deep Potential Attention (DPA) architecture~\cite{DPA1}.
To ensure uniform accuracy in evaluating the enthalpy landscape of ternary hydrides under experimental-achieved pressures of 150–250 GPa, the development of the LAM involved two concurrent learning iterations\cite{wang2024concurrent,dpgen} as illstrated in Fig.~\ref{fig:fig1}(b).
In the first stage, hydride structures with chemical formulas \ce{A_mB_nH_x} were generated using the CALYPSO structure prediction method, where A and B represent two distinct non-hydrogen elements, with $m \leq 4$, $n \leq 4$, and $3(m+n) \leq x \leq 10(m+n)$. 
These structures were then geometrically optimized using the L-BFGS algorithm, with an initial DPA model providing predictions for enthalpy and forces. 
Force prediction errors were evaluated along all relaxation trajectories, and configurations with significant errors were labeled using Density Functional Theory (DFT) calculations before being added to the training dataset.
This process was repeated iteratively until the force predictions for all configurations reached the desired accuracy.
The second stage followed a similar procedure but focused specifically on the enthalpy prediction accuracy of the relaxed structures.
Upon completion of the concurrent learning iterations, we compiled a dataset consisting of 218,349 samples, each labeled with energy, force, and stress values derived from DFT calculations.
The production model, trained on this dataset, demonstrated test accuracies of 37.5 meV/atom for energy predictions, 171 meV/Å for force predictions, and 45.7 meV/atom for virial predictions.
Further details on the learning procedure are provided in Supplementary Information Sec.~I. 

All potential chemical compositions of \ce{A_mB_nH_x} within the defined space, where \(m \leq 4\), \(n \leq 4\), and \(3(m+n) \leq x \leq 10(m+n)\), were systematically enumerated. 
For each composition, structure-searching calculations were then carried out at 200 GPa using the CALYPSO method, with the developed LAM serving as the computational workhorse to enable this large-scale exploration (Fig.~\ref{fig:fig1} (c)). 
A total of 36,292,030 ternary hydride structures were sampled, ensuring detailed coverage of the target chemical space.

Building on this extensive repository of compressed hydrides, a high-throughput screening workflow was established to identify the most promising candidates for high-temperature superconductivity (Fig.~\ref{fig:fig1}(d)). This workflow incorporates a series of filtering criteria, with a particular focus on fundamental properties concerning stability and superconducting behavior.  The filtering process is outlined as follows:

\begin{enumerate}
\item \textbf{Thermodynamic Stability (LAM Level)}: Thermodynamic stability was evaluated using the energy above the convex hull, calculated via LAM  ($E_{\hulllam}$). Hydrides with $E_{\hulllam} \leq 50$ meV/atom were retained, excluding those with unfavorable enthalpy. This reduced the initial dataset to 262,985 candidates.

\item \textbf{Symmetry Constraints}: To prioritize high-symmetry structures, only hydrides with a space group index above 75 were considered, focusing on orthogonal, trigonal, hexagonal, and cubic systems. This step reduced the pool to 9,984 candidates.

\item \textbf{Atomic Hydrogen Fraction}: Atomic hydrogen, a key factor in hydrides superconductivity, was quantified by identifying hydrogen atoms with interatomic distances exceeding 0.9 Å. Hydrides with an atomic hydrogen fraction greater than 75 \% were retained, narrowing the pool to 5,903 candidates.

\item \textbf{Dynamical Stability}: Dynamical stability was ensured by filtering out hydrides with imaginary phonon frequencies exceeding 3 THz, as predicted by the LAM. This step further reduced the candidate pool to 4,674 hydrides.

\item \textbf{Thermodynamic Stability (DFT Level)}: The thermodynamic stability of the remaining hydrides was re-evaluated at DFT level. Hydrides with $E_{\hulldft} \leq 50$ meV/atom were retained, leaving a final set of 2,429 candidates. (Supplementary Information Sec.~II for DFT computational details)

\end{enumerate}

This extensive high-throughput structure-searching effort, conducted on an unprecedented scale and powered by the LAM, guided by a physics-driven screening process, has yielded a 
repository of 2,429 candidate hydrides. 
These candidates demonstrate promising thermodynamic and dynamical stability, as well as significant potential for high-temperature superconductivity. This meticulously curated dataset provides a robust foundation for detailed electron-phonon coupling analyses, which are critical for the definitive identification of viable superconducting materials. 

However, performing comprehensive EPC calculations for all 2,429  candidate hydrides remains computationally prohibitive. To overcome this challenge, a two-stage EPC analysis was implemented to efficiently identify the most promising candidates. In the first stage, EPC calculations  (see Supplementary Information Sec.~III for further details) were performed for 480 out of the 2,429  hydrides, prioritizing systems with high symmetry, a low total number of atoms, and a high hydrogen-dominated density of states at fermi level. This process resulted in a total of 202 successful EPC calculations, yielding meaningful $T_c$ results after excluding structures with computational errors or imaginary phonon frequencies. Notably, among these 202 hydrides, 100 candidates were found to exhibit a calculated $ T_c > 200 \, \mathrm{K} $, distributed across 29 structural prototypes.

The second stage aims to extend the efficient search for high-\(T_c\) hydrides within the remaining pool of 1,949 candidates, selected from the initial 2,429. A \(T_c\)-prediction model is thus fine-tuned from the energy-pretrained LAM, utilizing data from the EPC calculations in the first stage  (see Supplementary Information Sec.~IV for further details). The $T_c$ model demonstrated a 4-fold test mean absolute error of 20.2$\pm$2.2K. 
The error of the model was reduced by 9.5~K (~30\%) compared to a model trained from scratch, thanks to the pretraining and multi-task fine-tuning strategy~\cite{zhang2024dpa}, which simultaneously fine-tunes the model on both the \(T_c\) dataset and the energy dataset, effectively mitigating catastrophic forgetting \cite{kirkpatrick2017overcoming}.
The remaining 1,949 hydrides were ranked based on their $E_{\hulldft}$ and predicted $T_c$ using the $T_c$ model, respectively.
The top-ranked candidates were subsequently subjected to detailed EPC calculations to further assess their potential as high-$T_c$ superconductors.
At this stage, 153 hydrides were analyzed, with 44 compounds in 14 structure prototypes identified as exhibiting a $T_c$ exceeding 200~K. 

The present theoretical framework, integrating high-throughput global structure searching, physics-guided screening, and first-principles-based EPC calculations, with the entire computational workflow enhanced by a deep-learning-based LAM, has collectively identified 144 potential high-temperature hydride superconductors across 38 distinct structural prototypes at 200 GPa, all with predicted $T_c$ exceeding 200 K. Importantly, all of them exhibiting good thermodynamic stability ($E_{\hulldft} \leq 50$ meV/atom). Of these, 129 hydrides and 27 structural prototypes represent novel predictions that, to the best of our knowledge, have not been previously reported in the literature. Previous extensive studies have identified about 30 structural prototypes for clathrate hydrides\cite{jiang2025data,duan_highthroughput}. In contrast, the present work introduces 27 previously unreported structural prototypes, representing a significant expansion—approximate doubling the number of known clathrate prototypes for high-temperature hydride superconductors.

\begin{figure}
\begin{center}
\includegraphics[width=0.50\textwidth]{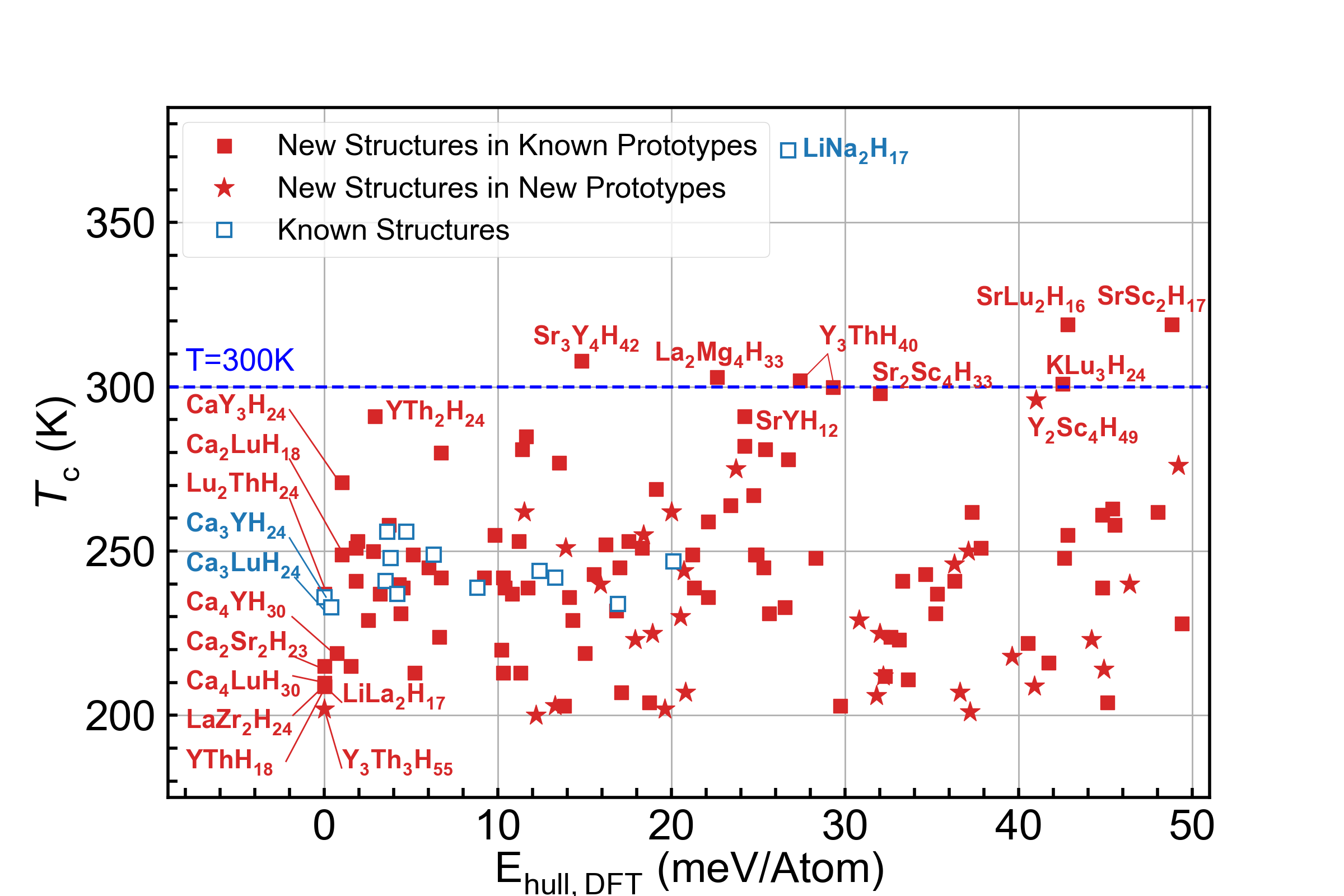}\\[5pt]  
\caption{Scatter plot of the 144 predicted ternary hydride superconductors in the $T_c$-$E_\hulldft$ space. Previously reported compounds reproduced in this study are represented by empty blue squares, while newly predicted compounds are shown as red circles. Compounds with novel structural prototypes are highlighted with red stars.} 
\label{fig:newfig2} 
\end{center}
\end{figure}

The scatter plot of the 144 predicted hydride superconductors in $T_c$-$E_{\hulldft}$ space is shown in Fig. \ref{fig:newfig2}. A complete list of these hydrides, categorized according to their structural prototypes, is provided in Supplementary Information Sec.~V. 12 hydrides are predicted to exhibit $T_c$ approaching or exceeding room temperature at a pressure of 200 GPa. These compounds include \hmn{R-3m}-\ce{Sr3Y4H42} ($T_c = 308$ K), \hmn{Fm-3m}-\ce{KLu3H24} ($T_c = 301$ K), \hmn{Fd-3m}-\ce{SrYH12} ($T_c = 291$ K), \hmn{Pm3m}-\ce{Y3ThH40} ($T_c = 302$ K), \hmn{R-3m}-\ce{Y3ThH40} ($T_c = 300$ K), \hmn{P6/mmm}-\ce{Y2ThH24} ($T_c = 291$ K), \hmn{Fd-3m}-\ce{SrLu2H16} ($T_c = 319$ K), \hmn{Fd-3m}-\ce{Li2NaH17} ($T_c = 372$ K), \hmn{Fd-3m}-\ce{SrSc2H17} ($T_c = 319$ K), \hmn{F-43m}-\ce{La2Mg4H33} ($T_c = 303$ K), \hmn{F-43m}-\ce{Sr2Sc4H33} ($T_c = 298$ K), and \hmn{P-6m2}-\ce{Y2Sc4H49} ($T_c = 296$ K). Notably, among these candidates, only \hmn{Fd-3m}-\ce{Li2NaH17} and \hmn{P6/mmm}-\ce{Y2ThH24} have been previously reported, underscoring the efficacy and predictive power of the present theoretical framework in discovering novel hydride superconductors.

To assess the synthesizability, we investigate thermodynamic stability of the 144 predicted hydride super-
conductors, and 8 hydrides are identified as lying on the convex hull, indicating that they are thermodynamically stable. These include \hmn{Fm-3m}-\ce{Ca3YH24} ($T_c = 236$ K), \hmn{R-3m}-\ce{Ca4LuH30} ($T_c = 210$ K), \hmn{P6/mmm}-\ce{Lu2ThH24} ($T_c = 237$ K), \hmn{P6/mmm}-\ce{LaZr2H24} ($T_c = 209$ K), \hmn{P-6m2}-\ce{YThH18} ($T_c = 209$ K), \hmn{Fd-3m}-\ce{Ca2Sr2H23} ($T_c = 215$ K), \hmn{I4/mmm}-\ce{LiLa2H17} ($T_c = 209$ K), and \hmn{R32}-\ce{Y3Th3H55} ($T_c = 202$ K). Among them, only \hmn{Fm-3m}-\ce{Ca3YH24} have been proviously proposed. It is noteworthy that that while the remaining hydrides are predicted to be metastable, they exhibit only marginal deviations from thermodynamic stability, all with $E_\hulldft$ less than 50 meV/atom, which also show the likelihood of their synthesis \cite{50meVref}.
Furthermore, these compounds could potentially be stabilized at pressures different from the current consideration. For instance, several hydrides in the Ca-Y-H, Ca-Lu-H, and Y-Lu-H systems are marginally metastable, with $E_\hulldft$ less than 10 meV/atom at 200 GPa, according to our calculations (Supplementary Information Sec. V). Previous studies suggest that these structures could be stabilized at higher pressures\cite{CaY3H24,CaYH12-258,YbLuH}. The \hmn{Fd-3m}-\ce{Li2NaH17}, which exhibits an exceptionally high $T_c$ of 372 K, has an $E_\mathrm{hull}$ of 26.7 meV/atom at 200 GPa, classifying it as metastable in the present work. However, this compound is predicted to become thermodynamically stable at higher pressures\cite{LiNaH}. Moreover, we indeed also did not consider other effects, such as anharmonic effect on these hydrides, due to computational demanding, where previous studies show that these effects may lower the stability pressure or affect superconductivity. Further investigations into the thermodynamic stability, as well as many other possible effects, including anharmonic effect and the exchange correlation function so forth, of the predicted compounds at different pressures are warranted for future studies. 

Within the chemical space investigated in this study, several ternary systems have been identified as hosting a substantial number of compounds that satisfy the established screening criteria, as summarized in Supplementary Information Sec. V. These systems emerge as strong candidates for prioritization in future experimental validation efforts due to their significant potential for the discovery of novel high-temperature superconducting hydrides. In particular, the Y-Lu-H, Ca-Y-H, Ca-Lu-H, Sr-Y-H, Na-Y-H, and Y-La-H systems predominantly feature hydrides with \ce{CaH6}-type structures. These structures are distinguished by variations in the occupation of A and B atoms within body-centered cubic (bcc) metal lattices. Moreover, this class of structures generally exhibits lower $E_\hulldft$ values compared to other structural prototypes, suggesting enhanced thermodynamic stability. Consequently, these systems are promising candidates for the experimental synthesis of nonstoichiometric alloy hydrides, as supported by prior experimental findings \cite{LaYH-experiment}. 

\begin{figure}
\begin{center}
\includegraphics[width=0.5\textwidth]{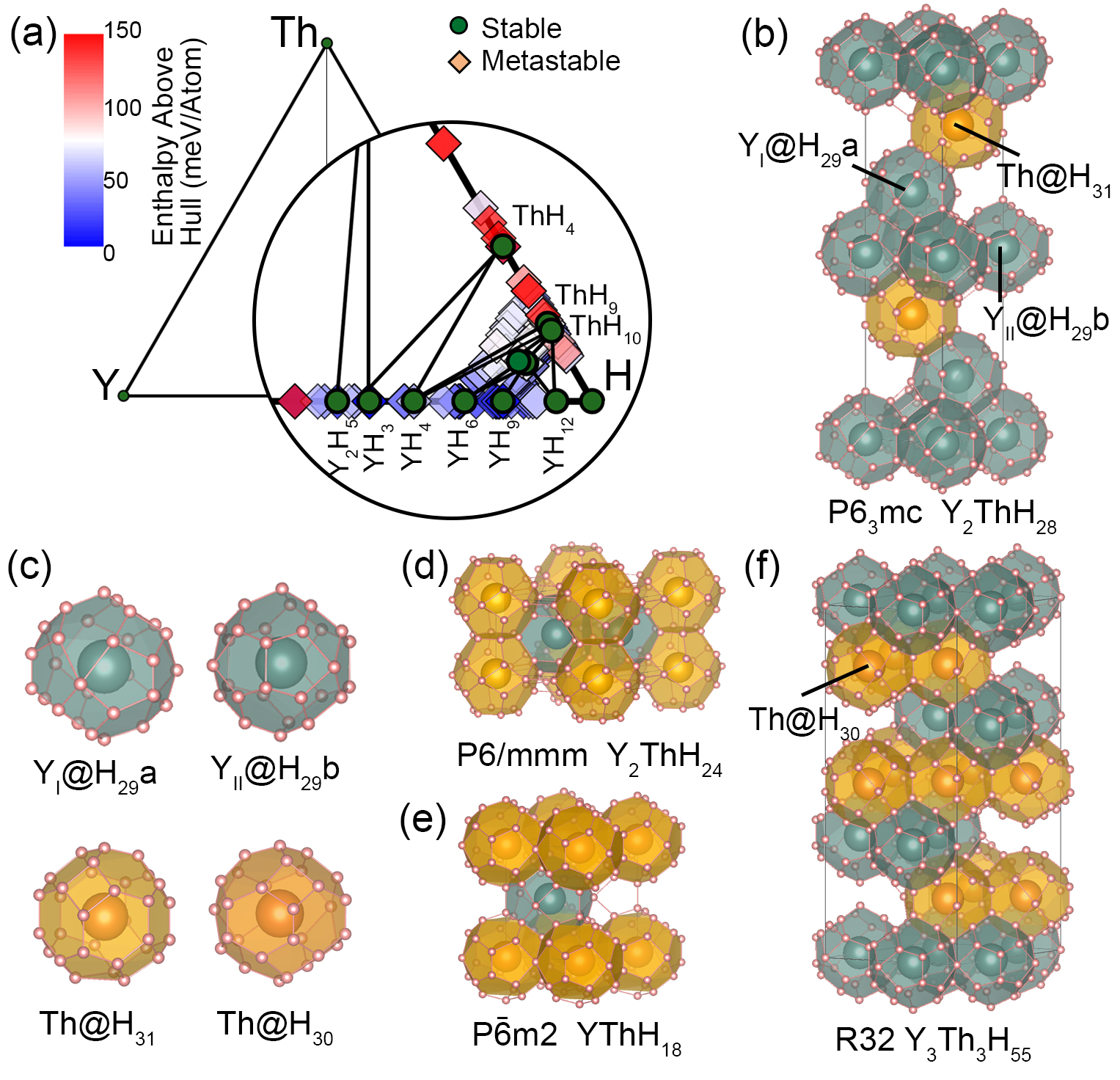}\\[5pt]  
\caption{The thermodynamic stability and crystal structure of representative hydrides in the Y-Th-H ternary system. (a) The ternary convex hull for the Y-Th-H system. (b) The crystal structure of  \hmn{P6_3/mc}-\ce{Y2ThH28}. (c) The detailed configurations of the two Y-centered \ce{H29} cages, namely \ce{H29}a and \ce{H29}b, and a Th centered \ce{H31} cages found in\ce{Y2ThH28}. \ce{H29}a and \ce{H31} cages are novel findings in \ce{Y2ThH28}, while \ce{H29}b is the typical \ce{H29} cage found in \ce{YH9} type structures. (d) The crystal structure of \hmn{P6/mmm}-\ce{Y2ThH24}, (e) \hmn{P-6m2}-\ce{YThH18}, (f) \hmn{R32}-\ce{Y3Th3H55}.} 
\label{fig:fig3}
\end{center}
\end{figure}

As a result of detailed analysis, the Y-Th-H system stands out as a particularly notable discovery, featuring 10 hydrides spanning 6 distinct structural prototypes that meet the screening criteria. Interestingly, structural prototypes in this system, apart from the conventional \ce{CaH6} and \ce{LaH10}, exhibit superior thermodynamic stability. The enthalpy convex hull for the Y-Th-H system at 200 GPa is presented in Fig.~\ref{fig:fig3} (a). Within this system, two metastable hydrides, \hmn{P6/mmm}-\ce{Y2ThH24} and \hmn{P6_3mc}-\ce{Y2ThH28}, are identified with $T_c \geq 250$~K. 
\hmn{P6_3mc}-\ce{Y2ThH28}, as shown in Fig.\ref{fig:fig3} (b), adopts a newly discovered structural prototype, featured by two types Y-centered \ce{H29} cages and a Th-centered \ce{H31} cage, as detailed in Fig.\ref{fig:fig3} (c). 
This compound with $E_\hulldft = 18.4 $ meV/atom  demonstrats a $T_c$ of 255 K at 200 GPa.
On the other hand, the \hmn{P6/mmm}-\ce{Y2ThH24}, as shown in Fig.\ref{fig:fig3} (d), shares the same structural prototype as the recently proposed \hmn{P6/mmm}-\ce{LaSc2H24} \cite{LaSc2H24-LHY} and exhibits a calculated $T_c = $ 291 K, approaching room temperature. 
This compound lies only 2.9 meV/atom above the convex hull at 200 GPa and is found to achieve thermodynamic stability at a lower pressure of 170 GPa\cite{jiang2025data}.
In addition, two thermodynamically stable compounds, \hmn{P-6m2}-\ce{YThH18} and \hmn{R32}-\ce{Y3Th3H55}, have been identified, with calculated $T_c$ of 209 K and 202 K, respectively. 
the \hmn{P-6m2}-\ce{YThH18} depicted in Fig.~\ref{fig:fig3}~(e) adopts the \ce{YH9}-type structure, in which Y and Th atoms are arranged in an hexagonal close-packed (hcp) lattice. 
In contrast, \hmn{R32}-\ce{Y3Th3H55} in Fig.~\ref{fig:fig3}~(f) represents a novel structural prototype that has not been previously reported. This unique structure consists of a network of Y-centered \ce{H29}, previously found in \ce{YH9}\cite{LaH10-theoretic2} and a new Th-centered \ce{H30} cage, as shown Fig.~\ref{fig:fig3}~(c). 
The presence of these structures demonstrates that the Y-Th-H system holds the great promise for the discovery of stoichiometric ternary hydrides with novel structural types and high-temperature superconductivity.

In summary, this study presents a deep-learning-driven computational framework designed to address the challenges of exploring the vast chemical and configurational space in materials discovery. 
By combining high-throughput global structure searching, physics-guided screening, and first-principles-based property calculations, enhanced by a deep-learning-based LAM, the framework serves as a powerful tool for accelerating the discovery of novel materials.
Applying this framework to ternary superhydrides resulted in the discovery of 129 compounds exhibiting a superconducting transition temperature greater than 200 K, with 10 of these compounds having a $T_c$ approaching or surpassing room temperature.
This investigation unveiled a diverse range of structural prototypes, encompassing both previously known clathrate-like configurations and entirely new structures, effectively doubling the catalog of high-$T_c$ hydride superconductors. 
These results underscore the predictive power and efficiency of the proposed approach in navigating the intricate landscape of multinary hydrides.

Future work will focus on extending this framework to quaternary and higher-order hydride systems, while systematically examining the effects of pressure variations on their stability and superconducting properties. Experimental validation will be crucial to confirm the predicted structures and $T_c$ values, as well as to provide deeper insights into the mechanisms underlying superconductivity. Overall, this study establishes a robust foundation for advancing high-temperature superconductivity, with far-reaching implications for both fundamental research and practical applications.

\bibliography{ref}{}
\bibliographystyle{apsrev4-2}




\end{document}